\documentclass{sig-alternate-05-2015}

\begin{document}

\setcopyright{acmcopyright}

\CopyrightYear{2016} 
\setcopyright{acmcopyright}
\conferenceinfo{CHASE'16,}{May 16 2016, Austin, TX, USA}
\isbn{978-1-4503-4155-4}\acmPrice{\$15.00}
\doi{http://dx.doi.org/10.1145/2897586.2897588}

\title{Useful Statistical Methods for Human Factors Research in Software Engineering: A Discussion on Validation with Quantitative Data}

\numberofauthors{2}
\author{
\alignauthor Lucas Gren \\
\affaddr Chalmers and University of Gothenburg \\
\affaddr Gothenburg, Sweden 412--92 and\\
\affaddr University of S\~ao Paulo\\
\affaddr S\~ao Paulo, Brazil 05508--090\\
\email {lucas.gren@cse.gu.se} \\
\alignauthor Alfredo Goldman\\
\affaddr University of S\~ao Paulo \\
\affaddr S\~ao Paulo, Brazil 05508--090\\
\email {gold@ime.usp.br} \\
}

\maketitle
\begin{abstract}
In this paper we describe the usefulness of statistical validation techniques for human factors survey research. We need to investigate a diversity of validity aspects when creating metrics in human factors research, and we argue that the statistical tests used in other fields to get support for reliability and construct validity in surveys, should also be applied to human factors research in software engineering more often. We also show briefly how such methods can be applied (Test-Retest, Cronbach's $\alpha$, and Exploratory Factor Analysis).  
\end{abstract}


%
%

%
%



\section{Introduction}
Science has from the beginning contributed enormously to the development of mankind. We have successfully observed the world and created models that help us understand and predict a diversity of events in the world, such as describing waves or the photoelectric effect. However, it is important to note that our predictive models are only models. As the famous statistician George E.P.\ Box said: 

\begin{quote}
``Essentially, all models are wrong, but some are useful.'' p. 424 \cite{box}. 
\end{quote}

The problem is that in more complex systems the deterministic models are no longer useful in the same way. This is when the mathematical models can be extended with probabilities. Stochastic models that express how likely an event is to occur then makes way more sense than setting out to describe all variables deterministically (which is often not feasible) \cite{stoch}.

The human mind is excellent at seeing patterns in a huge number of variables \cite{mind}. Therefore, when investigating human factors, it often makes sense to collect qualitative data and let the researchers (preferably, independently of each other) systematically look for patterns in the data set (e.g.\ a grounded theory approach) \cite{glaser}. However, as it is always good to look at a phenomenon (or construct) from different perspectives, a triangulation is always preferred. Therefore, it makes sense to collect quantitative data as well as qualitative and use statistical methods to analyze the former, i.e.\ using both words and numbers in the analysis \cite{qualmiles}.

Empirical software engineering (ESE) is a quite new research field compared to, for example, psychology. ESE has come a long way and made great advances, and some researchers have stressed the importance of evidence-based software engineering (see e.g.\ \cite{kitchenham2002preliminary}). They highlight an example were the same data was used for validation as for factor extraction, which of course should never be allowed. However, we believe the software engineering field is ready for having a more scientific and precise guidelines to quantitative survey data for human factors research. There seems to be a gap in the usage of statistical validation between the more technical aspects of software engineering and human factors aspects. 

When building prediction models for software, for example, a factor analysis seems to be recommended \cite{butte}, but in softer software engineering aspects it is not standard practice and many authors get survey studies published without such validation of scales (see e.g.\ \cite{chen2015perspectives,bosu2014peer,fernandez2015naming}). When it comes to human factors research in software engineering, it seems to be more depending on the statistical knowledge of the author then common practice for publication, i.e.\ many human factors survey studies have such tests in software engineering (see e.g.\ \cite{wohlin2004individual}), but far from all. 

Such methods are well-used in high impact management journals as well (see e.g.\ \cite{serrador,frohlich2001arcs}). The problem is that if one skips this part and directly run statistical tests on, for example, a measurement's correlation to another, they make little sense since we do not know if we managed to measure the intended construct. Skipping such validation would make editors of more mature fields reject the manuscript \cite{thompson1996factor}. A validation process is of course not only about statistical validation tests, but they should be conducted as an important step in the validation process \cite{exploratoryfactor1}. 

This paper will present some of these methods that have frequently been used in e.g.\ psychology for almost a century, directly applicable to human factors research in software engineering. This paper is organized as follows: Section~\ref{sec:human} describes the similarities between human factors research in software engineering and other more mature fields, Section~\ref{sec:methods} presents statistical test often used in the such fields, and Section~\ref{sec:disc} discusses the previous sections and their implications to software engineering research.

\section{Social Science Research}\label{sec:human}
Many subareas within software engineering have social science aspects. Many researchers have stated, for example, that agility is undeniably a soft as well as a hard issue \cite{doingtobeing,tolfo}. However, we prefer calling these types of issues simple (hard) and complex (soft) instead since these terms describe these issues better. The hard issue is a simple ditto with few and clear variables and the soft one is a complex adaptive social system.

In order to clearly explain how we believe this is applicable to software engineering we will use agile development processes as an example. Studies have been conducted that set out to investigate the social or cultural aspects of agile development. Withworth and Biddle \cite{whit}, for example, verifies that agile teams need to look at social-psychological aspects to fully understand how they function. There are also a set of studies connecting agile methods to organizational culture. These connect the agile adoption process to culture to see if there are cultural factors that could jeopardize the agile implementation, which there are \cite{iivari,tolfo2008}. One study divides culture in different layers depending on their visibility according to Schein \cite{scheincultlead}. That paper shows that an understanding of culture layers increase the understanding of how an agile culture could be established \cite{tolfo}. 

If we want to investigate these types of issues we can look at other fields that have been dealing with social science for more than half a century. If we use humans and their opinions in research we only investigate their perception of what is happening in the organization. Even if this is the case, we still need to check that items used to measure a construct manage to circle it somehow, i.e.\ items that are different but still correlated in relation to the construct under investigation. In order to do this we need two things; first, a reasonably large sample representative of the population (and large enough to remove individual and cohort bias), and second, make sure our items are correlated and pinpoint our construct of interest. The latter is often forgotten in software engineering research.

To simplify this explanation, let us look at a simple example. If a test (e.g.\ a survey) gives the correlation matrix in Table~\ref{fig:corr} the corresponding factor loadings would be the ones showed in Table~\ref{fig:factor}. 
\begin{table}
\caption{A correlation matrix.}
\centerline{\includegraphics[width=70mm]{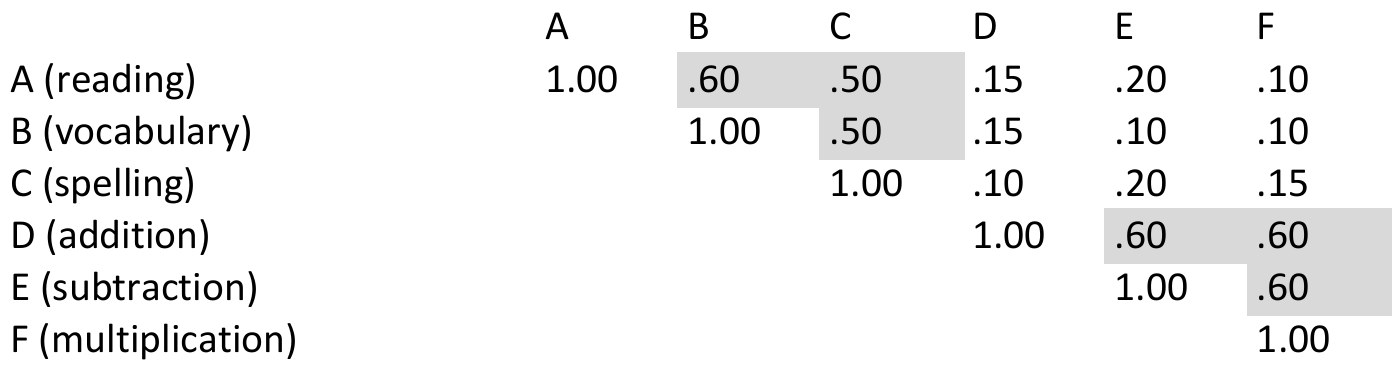}}

\label{fig:corr}
\end{table}

\begin{table}
\caption{The corresponding factor matrix with factor loadings.}
\centerline{\includegraphics[width=40mm]{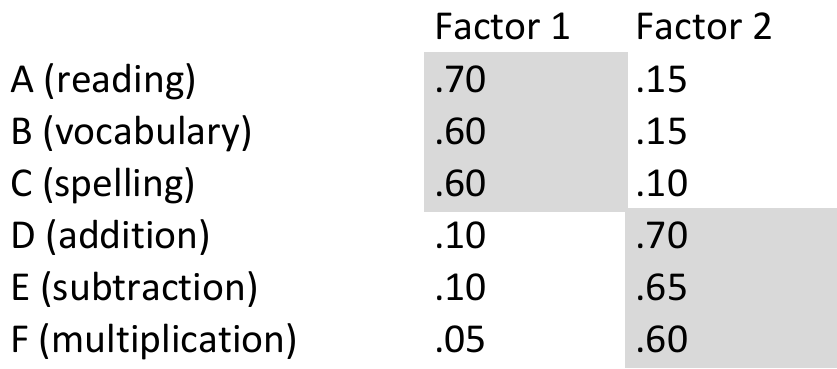}}

\label{fig:factor}
\end{table}

How to obtain the factors in a factor analysis is an advanced mathematical method and we will not go into details on how the calculations are conducted. However, the main reasoning behind the technique is that if $p$ is the number of variables $(X_{1},X_{2},...,X_{p})$ and $m$ is the number of underlying factors (the model presumes we have underlying factors) $F_{1},F_{2},...,F_{m}$ is that specific variable's representation in the latent factors. Each measured (or observed) variable is then a linear combination of the factors and reproduce maximum correlations: $X_{j} = a_{j1}F_{1} + a_{j2}F_{2} +,...,+ a_{jm}F_{m} + e_{j}$ where $j = 1,2,...,p$ and $a_{j1}$ is the factor loading of the $j^{th}$ variable on the first factor and so on. The factor loadings can be seen as weights (or contrast) of a linear regression model and tell us how much the variable has contributed to the factor. There are different extraction techniques for factor analysis and if the error variance is included it is called a Principle Component Analysis since we use all variance to find factors. If the error variance is excluded it is often called a Principle Axis Factor extraction, but the output pattern matrices are very similar \cite{fabrigar}. 

In this case we probably had reason to believe that there were two factors (or constructs) based on the variables in the study. If, for example addition and subtraction were uncorrelated in our result, we would have reason to doubt our measurement of the construct of mathematical skills. Even if a construct like ``mathematical skill''  is ambiguous, we still need to make sure our subset (like mathematical operators --addition, subtraction, and multiplication--) are valid. In this simple example we would have empirical support that items A, B, and C describe one construct and D, E, and F another, which also makes sense (i.e.\ high face validity). In Section~\ref{sec:methods} we explain this method in more detail. 

We can extend the same reasoning to that of ``agility''. Even if we do not have a clear definition of this term we can still research agile practices or behavior. For example, if we want to research Integration Testing or Retrospectives, we must use items that are different but correlate in a satisfactory way. The only more scientifically validated scale for agile practices that we found was the article ``Perceptive Agile Measurement:
New Instruments for Quantitative Studies in the Pursuit of the Social-Psychological Effect of Agile Practices by So and Scholl \cite{so}, where they use the method presented in this paper. As an example, their scale for ``Retrospectives'' (assessed on a Likert scale from 1 --Never-- to 7 --Always--) contains the items: (1) How often did you apply retrospectives? (2) All team members actively participated in gathering lessons learned in the retrospectives. (3) The retrospectives helped us become aware of what we did well in the past iteration\slash s. (4) The retrospectives helped us become aware of what we should improve in the upcoming iteration\slash s. (5) In the retrospectives we systematically assigned all important points for improvement to responsible individuals. (6) Our team followed up intensively on the progress of each improvement point elaborated in a retrospective.

These items were developed using two pretests and a validation study with a sample of $N$=227. When running such an analysis on other agile measurement tools, they often show problems with validity since the quantitative validation step was disregarded in its development \cite{grenjss}.

The whole discussion of ``if we measure what we think we measure'' is dealt with in most papers in the Validity Threats section. But what is validity? When it comes to tests in human systems we cannot just look at the measurement tool itself but also the context and interpretation of test scores, like in the definition of validity by Messick \cite{messick}:

\begin{quote}
``Validity is not a property of the test or assessment as such, but rather of the meaning of the test scores. These scores are a function not only of the items or stimulus conditions, but also of the persons responding as well as the context of the assessment. In particular, what needs to be valid is the meaning or interpretation of the score; as well as any implications for action that this meaning entails.''
\end{quote}
This means we always validate the usage of a test, and never the test itself. 

When soft issues are investigated in Psychology they are, of course, analyzed using both quantitative and qualitative data. However, after a more exploratory investigation (usually through qualitative case studies) we need to proceed and collect empirical evidence of the phenomenon (or construct) we found and see if numbers support our ideas. Of course we need a holistic view of validity and studies using quantitative data sometimes get undeserved credibility since the mathematical methods alone can seem advanced and serious. However, this is also what we see as a danger in software engineering survey research. If we create a survey and skip the statistical validation procedures we do not know what we measure. The statistical validation is only one aspect of the validation needed, but we should at least make sure that part supports our hypotheses.

\section{Useful Statistical Methods for Validation}\label{sec:methods}
There are many ways to categorize and list validity threats in different fields. In this paper we will only present a few aspect of validity where statistical tests can help us. These are: (1) Reliability: Stability. (2) Reliability: Internal Consistency. (3) Some aspects of Construct Validity.

The first in the list could quite easily be calculated by conducting a test and a retest in the same context, and then calculate the Pearson's correlation coefficient for the repeated measurements (which is a test for statistical dependence between two random variables or two data sets). If the first test is strongly correlated to the second (close to 1) we have some evidence that our test is stable. 

For testing the internal consistency of a scale, the most common method used is Cronbach's $\alpha$ \cite{cronbach}. Without providing a mathematical definition, the $\alpha$ can be seen as an overall correlation coefficient for a set of items. The $\alpha$ is a function of the number of items in a test, the mean covariance between item pairs, and the variance of the overall total score. The idea is that if a set of items is meant to measure a certain construct, the included items must be correlated. The Conbach's $\alpha$ can be used as a step in an Exploratory Factor Analysis (EFA), which is a way to test aspects of Construct Validity.  

The first step of a EFA is meant to investigate underlying variables in data (i.e. what factors explain most of the variance orthogonally). The next step is to rotate the factors and group them if they correlate and explain much of the same variance (i.e. the factors in a scale should not correlate too much or too little if they are considered to explain and measure a construct). A factor analysis is a statistical help to find groups of variables that explain certain construct in a data set. For more details, see e.g.\ \cite{fabrigar}.

The first thing to do when conducting a factor analysis is to make sure the items have the assumptions fulfilled for such a method, i.e.\ they need to be correlated to each other in a way that they can measure the same concept. Testing the Kaiser-Meyer-Olkin Measure of Sampling Adequacy and Bartlett's Test of Sphericity is a way to do this. Sphericity is how much the items (an overall test) are dependent on each other, i.e.\ the null hypothesis is that the items form an identity matrix (no correlations between items and therefore not suitable for a factor analysis). If this test is significant we can proceed with our analysis. The Kaiser-Meyer-Olkin Measure of Sampling Adequacy tests how correlated the items are, and if the value is under $<$.5 they are unacceptable as a rule of thumb and items with low correlations to the others need to be removed. Suitable items to remove can be selected based on such correlations from an Anti-Image table. If factors are removed and an acceptable value is obtained, we can create a Factor Matrix with our new suggested factors (again, for more details see e.g.\ \cite{fabrigar}). The extraction is often based on Eigenvalues $>$1 and the first factor will usually explain more variance than the following ones. This is because the algorithm tries to find a set of items that explain as much of the variance as possible, and then the second factors does that same thing for variance not explained by the first factor, and so on and so forth. 

We would also like to state that the factor rotation can be done based on the assumption that the items do are not dependent on each other (an orthogonal rotation) or that they could be dependent (an oblique rotation). In an oblique rotation we get information of how internally correlated the items in the different factors are. If the results are unsatisfactory, they could of course be used to reorganize factors and improve\slash rethink the included items. 

Getting ``enough'' data is always a tricky aspect of statistical tests, and maybe more so in software engineering than management or psychology. When it comes to factor analysis, the sample size needed is dependent on e.g. communalities between, and over-determination of, factors. Communality (or Internal Consistency) is the joint variables' possibility to explain variance in a factor, which can be calculated by the Cronbach's $\alpha$ mentioned above. Over-determination of factors is how many factors are included in each variable \cite{maccallum}.

\section{Discussion}\label{sec:disc}
To summarize, the method presented above is one way of checking if the numbers support the idea that a test really measures what we hope it does. This will not replace other aspects of validity but we do not see any disadvantages with collecting empirical evidence for surveys used in research. In the field of e.g.\ psychology, researchers need to be very careful stating that surveys, that have not been scientifically validated, give any kind of evidence for a certain research hypothesis. We believe the field of software engineering should be as careful when using poorly validated tools both in research and practice. 

Of course we realized that getting a lot of data can be cumbersome. Sometimes we need to do as well as we can given small samples and scarce information. However, the research community should be aware of the drawbacks and, at least, use statistical validation techniques when applicable to a data set. 

The main hinder for using such statistical methods is that the knowledge of statistics are generally lower than what is needed to understand the methods presented here. This is a dilemma in all social science research and we understand its implications, but we believe some more training in applied statistics and method (i.e.\ psychometrics) in software engineering education could somewhat reduce this knowledge gap. In addition, all tests we have presented here are implemented in most statistical software (including open source alternatives) and what is needed is to interpret the output. 

\bibliographystyle{abbrv}
\bibliography{references}  

\end{document}